\documentclass[%
 reprint,
 amsmath,amssymb,
 aps,
floatfix,
]{revtex4-2}

\usepackage{graphicx}
\usepackage{dcolumn}
\usepackage{bm}
\usepackage{xcolor}
\usepackage{textgreek}
\usepackage{gensymb}

\begin{document}

\preprint{APS/123-QED}

\title{High-frequency diode effect in superconducting Nb$_3$Sn
micro-bridges}

\author{Sara Chahid}
\affiliation{Advanced Physics Laboratory, Institute for Quantum Studies, Chapman University, Burtonsville, MD 20866, USA}

\author{Serafim Teknowijoyo}
\affiliation{Advanced Physics Laboratory, Institute for Quantum Studies, Chapman University, Burtonsville, MD 20866, USA}

\author{Iris Mowgood}
\affiliation{Advanced Physics Laboratory, Institute for Quantum Studies, Chapman University, Burtonsville, MD 20866, USA}

\author{Armen Gulian}
\email[Corresponding author: ]{gulian@chapman.edu}
\affiliation{Advanced Physics Laboratory, Institute for Quantum Studies, Chapman University, Burtonsville, MD 20866, USA}


\begin{abstract}

The superconducting diode effect has been recently reported in a variety of
systems and different symmetry breaking mechanisms have been examined. However, the 
frequency range of these potentially important devices still remains obscure. We
investigated superconducting micro-bridges of Nb$_{3}$Sn
in out-of-plane magnetic fields; 
optimum magnetic fields of $\sim$10 mT generate $\sim
$10\% diode efficiency, while higher fields of $\sim$15-20 mT quench
the effect. The diode changes its polarity with magnetic field reversal.
We documented superconductive diode rectification at
frequencies up to 100 kHz, the highest reported as of today.
Interestingly, the bridge resistance during diode operation reaches a value 
that is a factor of two smaller than in its normal state, which is compatible with the
vortex-caused mechanism of resistivity. 
This is confirmed by finite element modeling based on time-dependent 
Ginzburg-Landau equations. To explain experimental findings, 
no assumption of lattice thermal inequilibrium was required. 
Dissimilar edges of
the superconductor strip can be responsible for the inversion symmetry
breaking by vortex penetration barrier; visual evidence of this
opportunity was revealed by scanning electron microscopy. 
Estimates are in favor of much higher (GHz) range of frequencies for this type of diode. 

\end{abstract}

\maketitle


\section{Introduction}

For certain important problems of fundamental physics, such as the
exploration of possible quasi-local action of curl-less vector potential \cite{Gulian18}%
, it is necessary to explore superconducting micro-bridges for assurance that the
magnitudes of critical currents of the bridge are equal for both current 
polarities. In particular, accidental breaking of time reversal
symmetry (TRS) by remanent or spurious magnetic fields in the cryostat
together with simultaneous inversion symmetry (IS) breaking by, e.g., physical 
asymmetry of the bridge edges can cause the
superconducting diode effect (SDE) which violates this 
equality \cite{Tokura18,Wakatsuki18,Hoshino18,Ando20,Ideue20,BaumgartnerNN,Wu22,Strambini22}. 
While this equality violation should be avoided for the above-mentioned task, it is underlying for the SDE.

Recently, the SDE attracted noticeable attention within the research community 
\cite{Wakatsuki17,Wakatsuki18,Hoshino18,Tokura18,Ideue20,Ando20,Shin21,Strambini22,Wu22,BaumgartnerNN,BaumgartnerIOP,Bauriedl22,Yuan22,He22,Lin22,Ilic22,Karabassov22,Daido22,Anwar22,Souto22,Halterman22}, 
and the related activity generated multiple mechanisms applicable 
to understanding the SDE. In theoretical models, the TRS
is broken by an externally applied magnetic field or internal inclusions of
magnetic micro-clusters, while the IS is broken by the out-of-plane Rashba
spin--orbit coupling \cite{Yuan22,Ilic22,Daido22,Karabassov22,He22}, valley-Zeeman interaction \cite{Bauriedl22}, etc.,
which results in the emergence of a chiral superconducting order 
\cite{Yuan22,Ilic22,Daido22,He22}. Experimentally, systems based on van der Waals material $\mathrm{%
MoS}_{2}$ with noncentrosymmetric crystal potential \cite{Wakatsuki17}, synthetic
super lattice of $\mathrm{Nb/V/Ta}$ \cite{Ando20}, 
and intrinsically IS broken NbSe$_2$ \cite{Shin21} have
been reported as well as planar Josephson junction arrays of $\mathrm{Al}$
on $\mathrm{InAs}$ \cite{BaumgartnerNN,BaumgartnerIOP}. Yet other systems reveal nonreciprocal
behavior in field-free environments, such as $\mathrm{NbSe}_{2}$-based
Josephson junction \cite{Wu22}, tri-layer graphene \cite{Lin22}, and 
Josephson junction based on chiral superconductor Sr$_2$RuO$_4$ with the 
internally broken TRS \cite{Anwar22}. In view of
a variety of experimental observations and theories on the SDE in systems of different
configuration, it is quite reasonable to assume that more than
one mechanism can be responsible for the ubiquity of nonreciprocity observations
in superconductor thin films \cite{Hou22}. In a recent article \cite{Suri22}, the
SDE was found in $\mathrm{NbN}$ micro-bridges in an out-of-plane magnetic
field. This observation was attributed to the critical current being
determined by the vortex flow, confirming that the SDE is caused by unequal
vortex barriers on the two edges of the bridge \cite{Hope21,Vodolazov05}.

To understand this mechanism, consider a type-II superconductor film
strip. A magnetic field above a certain critical value creates vortices; initially, they
nucleate at the strip edges \cite{Shmidt1,Shmidt2}. Morphology of these edges affects
the surface barrier, which prevents the vortices from entering into the strip 
\cite{Bean64,FossheimBook}. However, if the current through the strip is
strong enough, the Lorentz force which it exerts onto the vortices overcomes
the surface barrier; vortices start moving across the strip, thus
dissipating energy and creating a resistive state. In practice, vortex
barriers are unequal because of the non-identical structure of strip edges.
This circumstance was used by Vodolazov and Peeters \cite{Vodolazov05} 
when predicting the SDE in 2005. It was 
experimentally observed recently \cite{Suri22}.

The spike of activities in the area of superconducting diodes paves the
way towards future practical application of these novel devices in
superconducting electronics. However, a very important topic is still open
and remains mainly unaddressed: the frequency range of the SDE. 
In the report by Lyu et al. \cite{Lyu21} experiments have been performed at 30 kHz;
however, the diode outputs were acquired by a DC nanovoltmeter, 
which leaves the dynamics of device obscure. Here, we directly detect the 
voltage response in the time domain at frequencies up to 100 kHz.
Experimental findings and modeling results provided us grounds to 
conclude that the SDE can perform at much higher frequencies.

\begin{figure*}[ht]
    \centering
    \includegraphics[width=\linewidth]{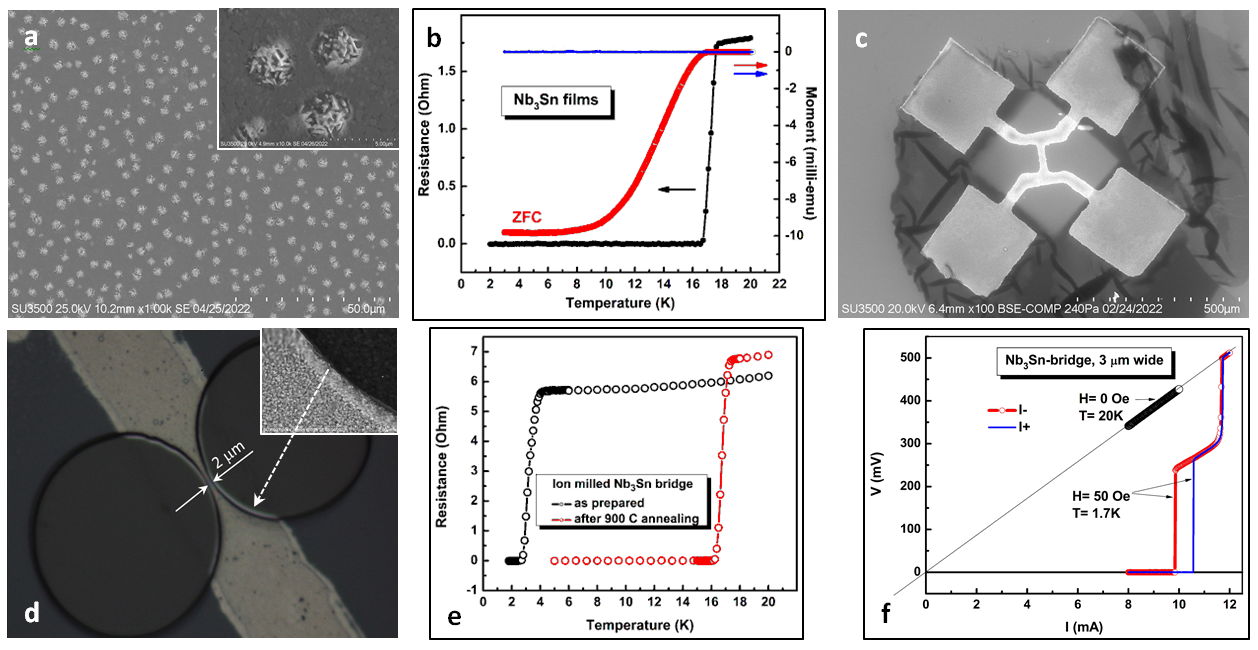}
    \caption{(\textbf{a}) Surface morphology of the films with
inclusion of small areas (nanomountains) with higher concentration of Sn
(zoomed tenfold in the inset). (\textbf{b}) Resistive and magnetic
transitions into superconducting state. (\textbf{c}) A bridge after ion
milling of 3D-printed pattern (initial stage of preparation). (\textbf{d})
Last stage of bridge preparation: the structure in panel (\textbf{c}) was
covered by positive photoresist, two circles were projected via
epi-fluorescent microscope and ion milled to reduce the active part of
bridges to micron-scale (the smallest, 2 \textmu m result is shown).
Inset illustrates micro-roughness of the edge line. (e) Effects of ion
milling and post annealing on transition temperature of bridges. (\textbf{f}%
) Typical voltage-current dependence of bridges above ($T=20$ $\mathrm{K}$)
and well-below ($T=1.7$ K) the superconducting transition.}
    \label{fig1}
\end{figure*}

\section{Experimental details}

Our study used conventional $\mathrm{Nb}_{3}\mathrm{Sn}$ superconducting
thin film bridges. The $\mathrm{Nb}_{3}\mathrm{Sn}$ films were prepared in a
DC/RF magnetron sputtering system (manufactured by AJA International, Inc.)
with a base pressure of $1 \times 10^{-8}$ Torr. The $\mathrm{Nb}$ target
(Kurt Lesker, $99.95\%$) was placed inside a DC gun while the $\mathrm{Sn}$
target (Kurt Lesker, $99.999\%$) was sputtered using an RF source to avoid
melting. The sapphire substrate (AdValue Technology, thickness $650$ \textmu
m, $C$-cut) was cleaned thoroughly with isopropyl alcohol before it
was mounted on the holder. In our chamber's configuration, the substrate
holder is at the center of the chamber facing upwards, while the (five) sputtering
guns are located at the top. The substrate is rotated in-plane
throughout the whole deposition process to ensure homogeneous deposition
layer over the whole surface. Our pre-deposition \textit{in-situ} cleaning
of the substrate involves heating it up to 900\textdegree C for 10 
min followed by a gentle bombardment of $\mathrm{Ar}^{+}$ ions at 400\textdegree C
for 5 min. The substrate was oriented to face the ion
gun squarely. For the $\mathrm{Nb}_{3}\mathrm{Sn}$ film deposition, we used
a ``stack+anneal" process similar to the
one in \cite{SundahlThesis}. Taking advantage of the Nb-Sn phase diagram 
\cite{Charlesworth70}, an alternating $\mathrm{Nb}$ and $\mathrm{Sn}$ stacks with
an excess $\mathrm{Sn}$ content (Nb:Sn ratio $\sim 2$) ``phase-locks" 
into the $\mathrm{Nb}_{3}\mathrm{Sn}$ composition when the excess Sn evaporates during
post-deposition annealing. For the first/base layer, $\mathrm{Nb}$ was
sputtered in DC mode with 500 W power and 3 mTorr
pressure at 600\textdegree C for 20 min. Then, the Sn
layer was deposited on top of $\mathrm{Nb}$ in RF mode with 250 W
power and 30 mTorr pressure at 100\textdegree C for 15 
min. Nb was deposited one more time as the capping (third) layer
using the same parameters as the base layer for 3 min. Finally,
the substrate was heated \textit{in-situ} to 950\textdegree C for 30
minutes to let the excess Sn evaporate before cooling down to room
temperature. All the heating/cooling protocols consistently used a 
30\textdegree C/min ramp rate. Our films are $\sim 100$ nm thick
and resistivity measurements in PPMS (Quantum Design) indicated the correct Nb$_3$Sn
stoichiometry and homogeneity in view of their consistently measured
$T_{c}=17.6$ K and RRR($300$K/$20$K) $\sim 4$ between
samples across several batches, which was corroborated by VSM measurements.
Lastly, the stoichiometry was also checked by the
SEM EDX (Hitachi SU3500, Oxford Instruments X-MAX-20).

Figure 1 demonstrates
certain physical properties of our films (panels \textbf{a} and \textbf{b})
and bridges (panels \textbf{c}-\textbf{f}). 

\begin{figure}
    \centering
    \includegraphics[width=0.75\linewidth]{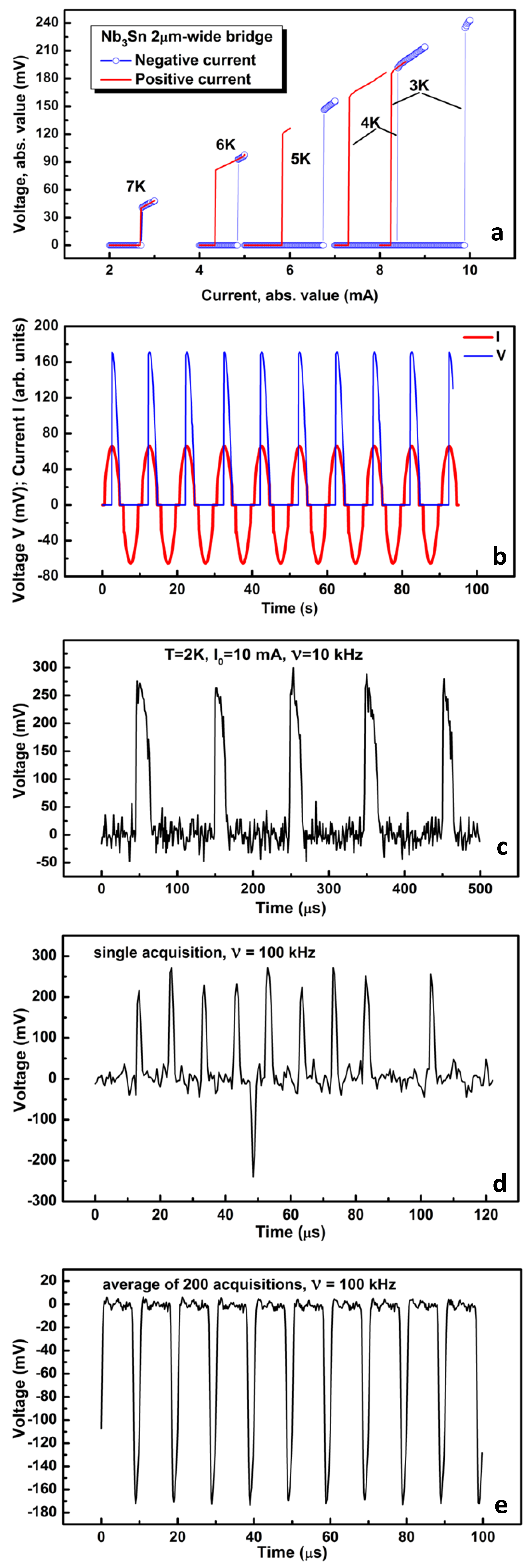}
    \caption{(\textbf{a}) Resistive state at various values of bias
temperature for the applied current of opposite polarities (absolute values are plotted). (\textbf{b}) The SDE
in 2 \textmu m bridge at sinusoidal current amplitude $\approx 10$ mA and frequency 0.1
 Hz. (\textbf{c}) Same as
in panel (\textbf{b}) with 5 \textmu m-wide bridge at frequency 10 kHz. (\textbf{d}) 
Same as in panel (\textbf{c}) at
frequency 100 kHz. (\textbf{e}) Average of 200 acquisitions with reversed
polarity of magnetic field. In all cases, a magnetic field orthogonal to the surface of the bridge was applied and optimized in the range 50-100 Oe.}
    \label{fig2}
\end{figure}

Transformation of the films into bridges was performed using a combination of 
3D-printing (Elegoo Mars-3 printer), photolithography and ion milling to achieve
macroscopic contact pads on $1\times 1$ cm$^2$ sapphire
substrates. Reactive ion milling (Bal-Tec RES-101, $\mathrm{CF}_{4}$
etchant) was used to develop a metallic film pattern, Fig.~\ref{fig1}(\textbf{c}).
After removal of the resin, the structure was covered by negative
photoresist (PKP-308PI, Transene Co., Inc.), and the 100 \textmu m-scale bridge was narrowed down
to 12 \textmu m with the projective photolithography (using mask
projection via LUMAM epi-fluorescent microscope, $10\times $ objective).
After ion milling, the last stage of patterning was undertaken using
positive photoresist (TRANSIST PC800, Transene Co., Inc.) and the same projective technique with the $40\times $
objective and a different mask in the form of two holes. The final ion
milling delivered bridges about 10-20 \textmu m long and down
to 2 \textmu m wide, Fig.~\ref{fig1}(\textbf{d}).

Ion milling affects the physical properties of bridges thus reducing the
critical temperatures down to the 2-12 K range. This $T_{c}$%
-reduction was also noticed in the literature \cite{Suri22}. We were able to restore
the $T_{c}$ values up to 17 K in some of the bridges by
high-vacuum post-annealing ($\sim 10^{-6}$ Torr) at 
900\textdegree C for 30 min, as shown in Fig.~\ref{fig1}(\textbf{e}).

\section{Results}

The tests demonstrated $|V(I_{+})|\neq |V(I_{-})|$, Fig.~1(\textbf{f}), thus indicating
the possibility of an SDE. At various temperatures, this type of the 
$V(I)$-test was systematically performed on a 2 \textmu m bridge 
and the results are shown in Fig.~\ref{fig2}(\textbf{a}).
These measurements confirm the fact mentioned in literature that at lower
temperatures the difference $\Delta I$ between the threshold values of $I_{+}^{res}$\
and $I_{-}^{res}$\ increases with the decreasing bias temperature \cite{Ando20,Hou22} (the
value of $I^{res}$ corresponds to the current at which the resistive state
emerges).

To confirm the expected diode effect, the bridge then was biased at $2$ K (PPMS
DynaCool cryostat) with external current source and nanovoltmeter
(Keithley 6221 and 2182a respectively) applied at frequency 0.1 Hz.
Figure~\ref{fig2}\textbf{b} demonstrates the result optimized at magnetic field 75 Oe.

Since superconducting diodes are considered to be important elements for
electronics, for the majority of applications, it is meaningful to register the effect at possibly higher
frequencies. The Keithley current source we used generates AC-currents up to 100 
kHz. Corresponding detection of voltage output was performed by an oscilloscope
(Tektronix TDS 644A). The results are shown in Fig.~\ref{fig2}(\textbf{c}) and (\textbf{d})
for 10 kHz and 100 kHz frequencies, respectively. Higher frequency
measurements require specially-designed circuitry in the cryostat since 
noise becomes an issue, as is seen from the comparison of 
panel (\textbf{b}) with panels (\textbf{c}) and (\textbf{d}) in Fig.~2. 
For noise reduction, we plotted 100 kHz
data averaged over 200 acquisitions, Fig.~2(\textbf{e}).

\section{Discussion}

\begin{figure}
    \centering
    \includegraphics[width=\linewidth]{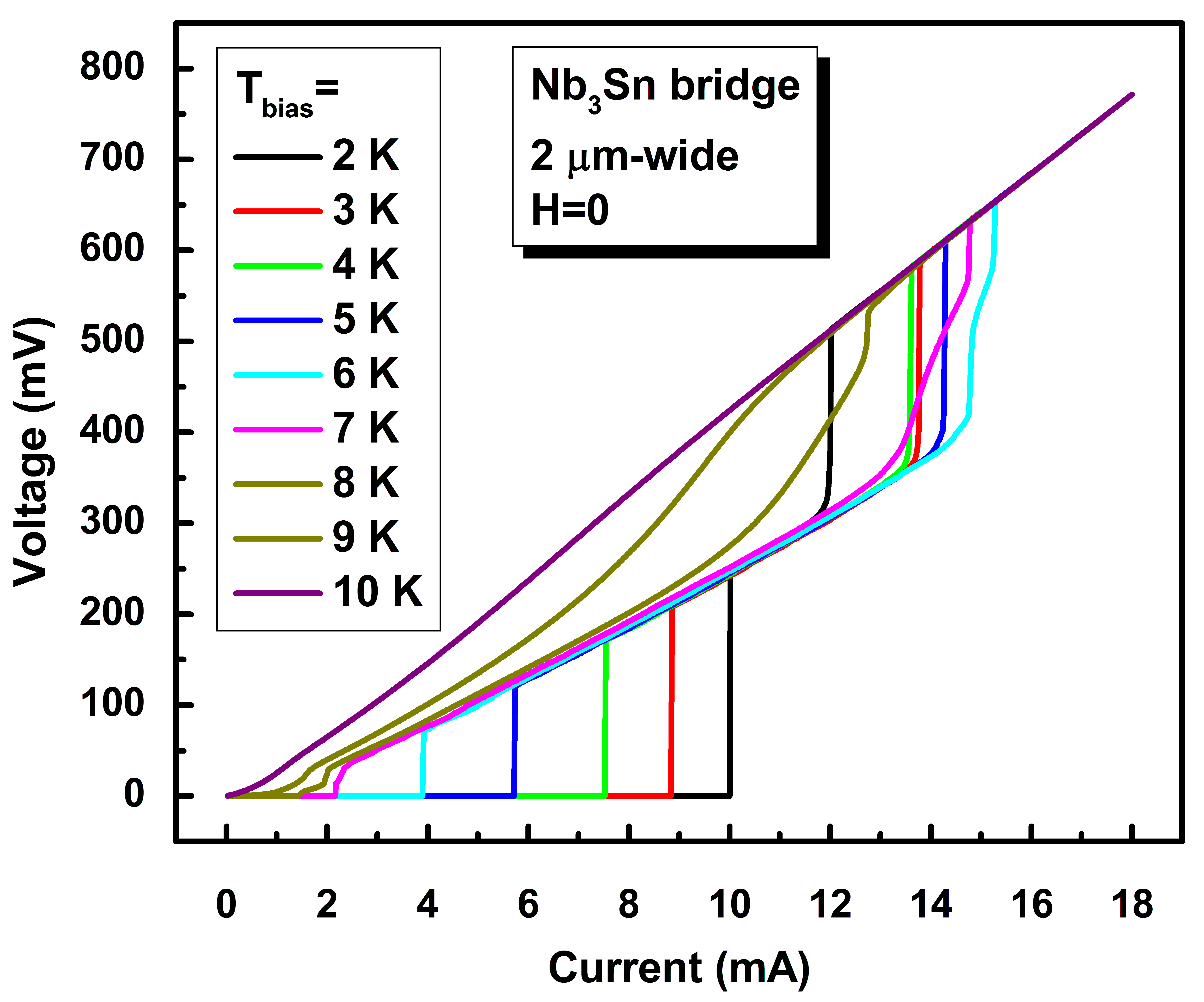}
    \caption{$V(I)$ dependence of 2 \textmu m-wide bridge. Resistive state jumps are noticeable
before the transition to the normal state as well as at the intermediate current.}
    \label{fig3}
\end{figure}

To estimate the frequency range of this type of diode, one needs to
understand what the underlying mechanism of the SDE is. 
As was expected from the general reasons mentioned in Introduction, 
the effect is nonexistent in the absence of external magnetic field $\mathbf{H}_{ext}$.
Comparison of panels
(\textbf{d}) and (\textbf{e}) in Fig. 2 shows the change of diode polarity 
due to the reversal of the external magnetic field from 100 Oe to 
-100 {Oe}. The field $\mathbf{H}_{ext}$ is orthogonal to the
bridge surface. So is the vector of internal
magnetic field, $\mathbf{H}_{int}$, though only at the lateral (top and bottom) edges of the bridge. 
This field is generated by the current flow through the bridge, with the field lines circulating it. Thus, from symmetry considerations, for an infinitesimally thin bridge, it is orthogonal to the bridge surface at the top and bottom edges. At one edge $\mathbf{H}_{int}$ is parallel, and on the other edge - antiparallel to $\mathbf{H}_{ext}$. Thus, on one of these
lateral edges, the amplitudes of external and internal fields add up
while on the other edge they subtract. 
In the AC case, this adding and
subtracting reciprocate periodically with frequency $\omega$: 
\begin{equation}
B=\begin{cases}
H_{ext}+ H_{int}\sin (\omega t) & \text{for the top edge}\\
H_{ext}- H_{int}\sin (\omega t) & \text{for the bottom edge}.
\end{cases}
\label{1}
\end{equation}
The $\mathbf{H}_{ext}-$field amplitude of $\sim 100$ Oe is optimal for
the SDE in our bridges. Higher values ($\sim 150$ Oe) quench the
effect. The resistive state which yields a non-zero voltage in Fig. 2(%
\textbf{c}) is caused by vortex motion: the corresponding value of the
resistance is a fraction of its normal value. This strongly distinguishes our SDE
from the via conformal-mapped nanoholes SDE \cite{Lyu21}, where the normal
state transition stands for the resistance, and flux pinning on defects
(nanoholes) secures the lossless current flow, and also breaks the IS. In
our case, as follows from Fig. 2(\textbf{c}), the resistance $(V/I)$ reaches
values of 25-40 $\Omega$. At the same time, full transition to the
normal state corresponds to the resistance of $\sim 60$ $\Omega$. Figure 3
depicts a set of $V(I)$ measurements at various temperatures. Multiple
resistivity jumps are noticeable at lower temperature curves. This step-wise
behavior of the $V(I)$ curves is 
explained via vortex pattern rearrangement \cite{Berdiyorov09,Lotero14,Vodolazov07}. Interestingly, Berdiyorov et al. \cite{Berdiyorov09} predicted a non-monotonic behavior of critical current of transition from resistive flux-flow state into normal state (calling this current $j_{c3}$) for an increase in magnetic field. In our case, a non-monotonic behavior of $j_{c3}$ is observable with the bias temperature. This fact requires explanation via modeling; however, it is beyond the scope of this report.

To explain the results shown in Fig. 2, modeling is necessary which, 
unlike \cite{Lyu21}, does not involve a transition into the normal state for resistivity.
Using COMSOL Multiphysics, we designed a $p-$parameter model \cite{GulianCOMSOL} in which the IS is
geometrically broken: the top edge of the strip has a weaker barrier to
vortex penetration than the bottom edge. One can assume, as in the case of 
NbN SDE \cite{Suri22}, that the IS is broken by opposite edges
asymmetry caused by physical reasons during the bridge preparation. A closer look
at Fig. 1 (\textbf{d}, inset) illustrates the fractal structure of the edge
of the current-flow channel. Realistically, the opposite edges are not 
microscopically symmetric, so our assumption looks quite plausible.

Prior to discussing the details of modeling, we should make one more
important remark. Sapphire, which we used as a substrate material, 
has a very high heat conductivity at low temperatures - much higher than Si, 
which was used in \cite{Lyu21}. This allowed us when modeling to decouple the dynamics of
superconducting electrons from the thermal deviation of the lattice from
equilibrium. The lattice just serves as an equilibrium heat sink at some
effective temperature $T$. The system of time-dependent Ginzburg-Landau
(TDGL) equations for the Cooper-pair wave function $\Psi =\mathrm{Re}\Psi +i%
\mathrm{Im}\Psi =\psi _{1}+i\psi _{2}$ in the dimensionless form can be
presented as (see, e.g., \cite{GulianCOMSOL}): 
\begin{equation}
\begin{split}
\dot{\psi}_{1}=&\frac{1}{\kappa ^{2}}\left( \psi _{1.xx}+\psi _{1.yy}\right) +%
\frac{2}{\kappa }\left( A_{1}\psi _{2.x}+A_{2}\psi _{2.y}\right) \\
+&\frac{\psi
_{2}}{\kappa }\left( A_{1.x}+A_{2.y}\right) -\psi _{1}\left(
A_{1}^{2}+A_{2}^{2}\right) \\
+&\psi _{1}\left( 1-\psi _{1}^{2}-\psi
_{2}^{2}+p\right)   \label{2} 
\end{split}
\end{equation}%
\begin{equation}
\begin{split}
\dot{\psi}_{2}=&\frac{1}{\kappa ^{2}}\left( \psi _{2.xx}+\psi _{2.yy}\right) -%
\frac{2}{\kappa }\left( A_{1}\psi _{1.x}+A_{2}\psi _{1.y}\right) \\
-&\frac{\psi
_{1}}{\kappa }\left( A_{1.x}+A_{2.y}\right) -\psi _{2}\left(
A_{1}^{2}+A_{2}^{2}\right) \\
+&\psi _{2}\left( 1-\psi _{1}^{2}-\psi
_{2}^{2}+p\right)   \label{3}    
\end{split}
\end{equation}%
\begin{equation}
\begin{split}
\sigma \dot{A}_{1}=-&\frac{1}{\kappa }\left( \psi _{2}\psi _{1.x}-\psi
_{1}\psi _{2.x}\right) -\left( \psi _{1}^{2}+\psi _{2}^{2}\right)
A_{1}\\
+&A_{1.yy}-A_{2.xy}  \label{4}    
\end{split}
\end{equation}%
\begin{equation}
\begin{split}
\sigma \dot{A}_{2}=-&\frac{1}{\kappa }\left( \psi _{2}\psi _{1.y}-\psi
_{1}\psi _{2.y}\right) -\left( \psi _{1}^{2}+\psi _{2}^{2}\right)
A_{2} \\
+&A_{2.xx}-A_{1.xy}  \label{5}    
\end{split}
\end{equation}%
Here $\kappa =\lambda _{L}(T)/\xi (T)$ is the Ginzburg-Landau parameter, $%
\sigma =0.172$ is the dimensionless conductivity, $\mathbf{A}=\mathbf{\hat{x}%
}A_{1}+\mathbf{\hat{y}}A_{2}$ is the vector potential (we chose the unit
vector $\mathbf{\hat{x}}$\ to be along the transport current flow in the strip,
with $\mathbf{\hat{y}}$\ orthogonal to it). The parameter $p=p(x,y)$, if
negative, generates local ``weakening" of the superconducting order parameter
\cite{GulianCOMSOL} and, consequently, breaking of the IS. In our model we chose $p=p(y)=-0.3$ along the top
edge within a $\lambda _{L}(T)$ distance from that edge. Since both
superconducting and normal currents are absent across the lateral edges of
the strip, we imposed zero-flux conditions for $\partial \Psi /\partial 
\mathbf{n}$ on these boundaries and periodic boundary conditions on
transverse facets (i.e., on the boundaries orthogonal to the current flow).
Initial conditions for the $\Psi -$function should not be identical zero;
for example, $\psi _{1}(t=0)=1$ and $\psi _{2}(t=0)=0$ may be chosen.
Boundary conditions for the $\mathbf{\hat{y}-}$component of the vector
potential $\mathbf{A}$ are also periodic on the transverse facets of the
strip. On the lateral edges they are $\mathrm{curl}\mathbf{A}=\mathbf{B},$
which is implemented via the flux-source conditions. Electric field in the
chosen gauge for Eqs. (2)-(5) with the zero scalar potential, $\varphi \equiv 0$, is $\mathbf{%
E}=-\partial \mathbf{A}/\partial t$, which means that the solution $\mathbf{A%
}=\mathbf{A}(t)$ is an explicit function of time. As soon as this function
is known in the whole $2D-$volume of the strip, one can average $A_{1.t}$
over this volume, and determine, up to a constant, the voltage $V$ caused by
the transport current between the transverse facets of the strip.

The dimensionless spatial distance in the adopted system of equations
corresponds to the London penetration depth, $\lambda _{L}(T)$, while the
dimensionless unit of time $t_{0}=\pi \hbar /\left[ 8k_{B}\left( T_{c}-T\right) \right] $
has picosecond order of magnitude if $T_c \sim 17$ K and $T \sim 2$ K. 

\begin{figure}
    \centering
    \includegraphics[width=\linewidth]{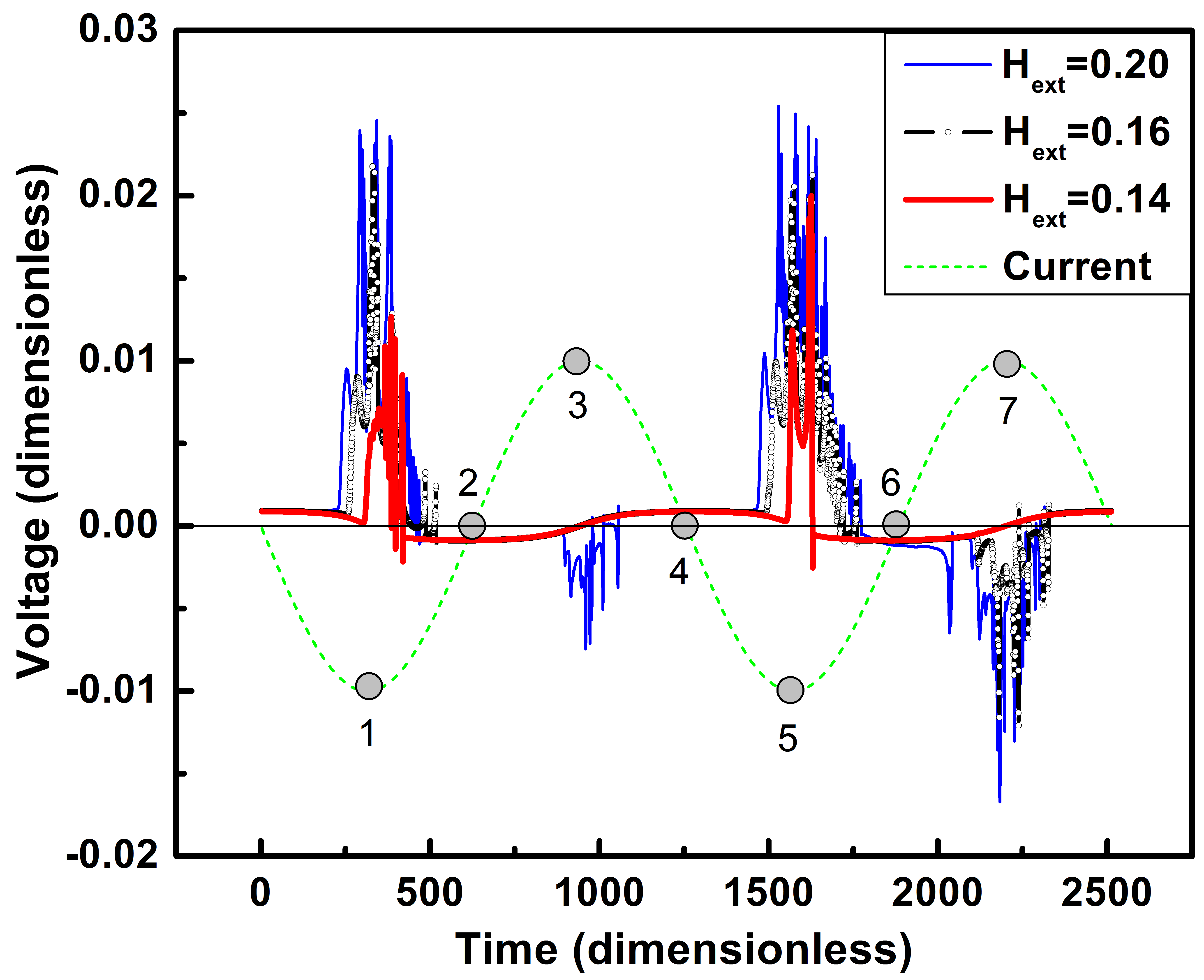}
    \caption{Voltage across the diode (superconducting bridge) generated by a 
sinusoidal AC bias: $j=j_0$ sin($\omega t+\pi$), where $\pi$ is added for computational 
convenience. The amplitude of $j_0$ is chosen to yield $H_{int}=-0.48$. Different 
curves correspond to different values of external magnetic field $H_{ext}$.}
    \label{fig4}
\end{figure}

The results of the modeling are shown in Fig. 4.
Figure 5 explains the physical mechanism behind the behavior of the voltage in 
Fig. 4. A video file in the \textit{Online Supplemental Materials} 
provides more details related to both of these figures.

\begin{figure}[h!]
    \centering
    \includegraphics[width=0.95\linewidth]{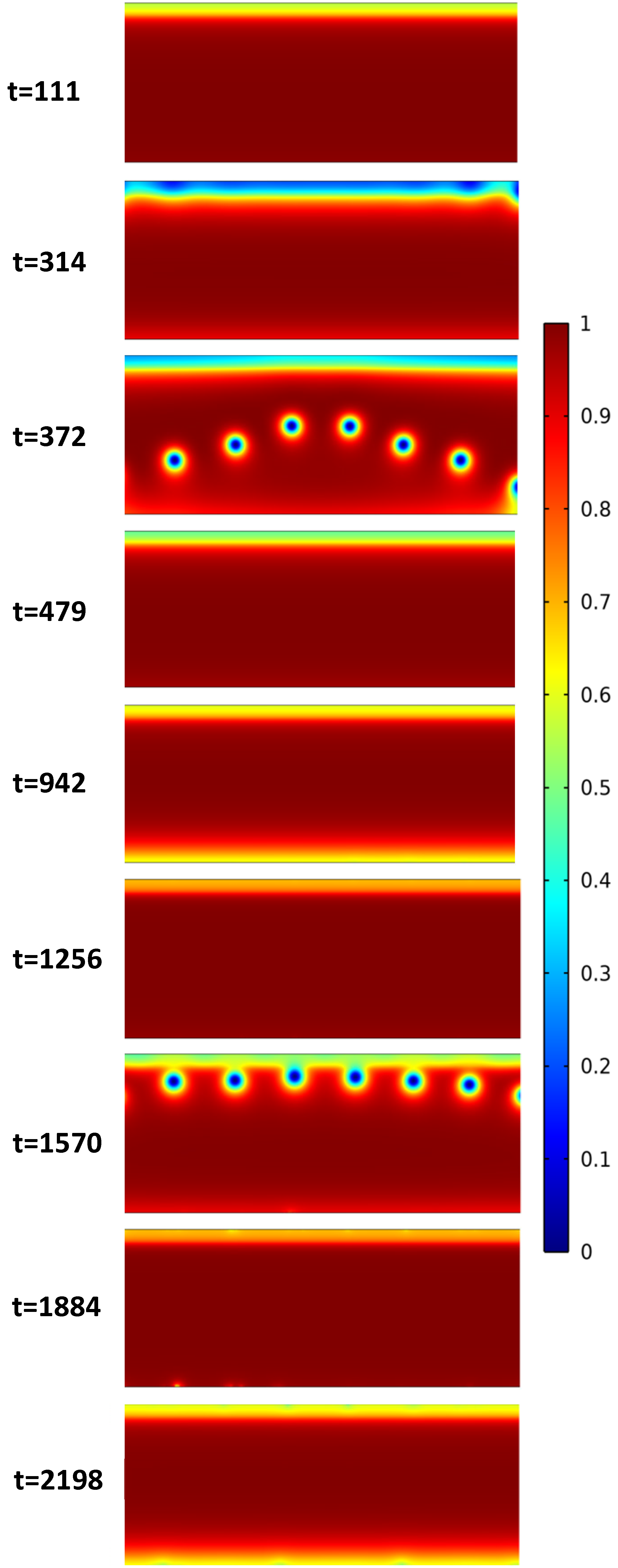}
    \caption{Physical processes in the active part of superconducting
bridge resulting in the SDE for $H_{ext}=0.14$, $H_{int}=-0.48$.}
    \label{fig5}
\end{figure}

As follows from Fig. 5, for the initial segment of periodic current (say, at $t=111$)
no fluxons are in the bridge. The current flow is lossless ($V\approx 0$),
and the Cooper pair density is homogeneous within the bridge except the top edge,
where it is weakly suppressed due to the asymmetry caused by 
the $p-$parameter, mimicking the IS break.
In accordance to Eq. (1), the magnetic field in the top
edge of the bridge at this time segment is higher than at the bottom. Tuning
the amplitude of the current, or/and the value of external field amplitude $H_{ext}$, one can exceed
the critical value of $\mathbf{B}$ at the top edge. This allows vortices 
to start penetrating from the top edge into the bulk of the
bridge. The relative values of $H_{ext}$ and $H_{int}$ should be chosen such
that the Lorentz force will be strong enough to move these vortices
across the bridge during the first half-period of $\sin (\omega t)$
function. In the second half-period, the value of $\mathbf{B}$ is larger at
the bottom edge. However, because the ``rigidity" of this edge 
(i.e., the Bean-Livingston barrier) is higher, the
value of $\mathbf{B}$, which was critical for the top edge, is not enough 
to also be critical for the bottom edge. So even at $t=940$, when $\mathbf{B}$
is maximal at the bottom, no vortices penetrate from the bottom
(just the density of Cooper pairs is changed a bit). This evolution 
displayed in Fig. 5 occurs at $H_{ext}=0.14$ (and $H_{int}=-0.48$). If
these values are correctly chosen, during the next period the process
repeats. In case of improper choices (e.g., $H_{ext}=0.16$ or $H_{ext}=0.20$ with the same $H_{int}=-0.48$), 
diode malfunctioning can occur. This kind of behavior was also noticed
during the experiments (see Fig. 2\textbf{d}). 

Associating the SDE with the described vortex motion, we can make certain estimates
for the performance limits of the SDE in the frequency domain. In the modeled
example, the width of the bridge should constitute units of $\lambda _{L}$. The angular
frequency was chosen to be $\omega =0.005$. It corresponds to frequency $%
\nu \sim$ 300 MHz. This value is in accordance with our
experimental observations. The 100 kHz results have been obtained
for a $5$ \textmu m-wide bridge. If the geometry is scaled down from 5
\textmu m to 5-50 nm-wide bridges, it can deliver frequencies 2-3 orders 
of magnitude  higher (10-100 MHz range) because of geometric factors
affecting their escape time. One can ask: 
can an even higher frequency range be reached? The answer appears to be yes, 
due to the following reason: the speed of kinematic vortices is 3 orders of
magnitude higher than that of the Abrikosov vortices \cite{Weber91,Andronov93}. If the $%
I_{+/-}^{res}$ asymmetry enters the range of kinematic
vortices (the highest current $V(I)$ jump area in Fig. 3 which corresponds 
to the transition into normal state), then the operational
frequencies of superconducting diodes can reach 100 GHz. 
Obviously, more research is required to reach that stage
experimentally and more modeling is required for better guidance on how to reach
this regime.

\section{Summary}

We successfully demonstrated the SDE in ordinary thin superconducting 
\textmu m-size bridges of Nb$_3${Sn} with $%
T_{c}=17$ K. The TRS was broken by external magnetic fields less
than 100 Oe, and the IS was broken by structural asymmetry between the
bridge edges. High-frequency (up to 100 kHz) diode performance
was documented in the time domain and evidence regarding nonequilibrium flux
quanta dynamic mechanism of diode action was obtained. Using this evidence, 
a qualitative explanation of the diode action was obtained via finite element
modeling using TDGL equations. Based on both our experimental and modeling results, 
it is possible to conclude that this kind of SDE can be functional at 2-3 orders
higher frequencies than the upper limit of our instrument (100 kHz). 
With the involvement of kinematic vortices into the diode 
action, an even higher range of frequencies (of the order of 100 GHz) can be achieved.
More experiments guided by modeling can yield two-terminal and multi-terminal
high-performance devices for superconducting electronics.

\begin{acknowledgments}

The work of Chapman U. research team is supported by the ONR grants N00014-21-1-2879 and
N00014-20-1-2442. We are grateful to Physics Art Frontiers for the provided technical assistance.

\end{acknowledgments}





\begin{thebibliography}{39}%
\makeatletter
\providecommand \@ifxundefined [1]{%
 \@ifx{#1\undefined}
}%
\providecommand \@ifnum [1]{%
 \ifnum #1\expandafter \@firstoftwo
 \else \expandafter \@secondoftwo
 \fi
}%
\providecommand \@ifx [1]{%
 \ifx #1\expandafter \@firstoftwo
 \else \expandafter \@secondoftwo
 \fi
}%
\providecommand \natexlab [1]{#1}%
\providecommand \enquote  [1]{``#1''}%
\providecommand \bibnamefont  [1]{#1}%
\providecommand \bibfnamefont [1]{#1}%
\providecommand \citenamefont [1]{#1}%
\providecommand \href@noop [0]{\@secondoftwo}%
\providecommand \href [0]{\begingroup \@sanitize@url \@href}%
\providecommand \@href[1]{\@@startlink{#1}\@@href}%
\providecommand \@@href[1]{\endgroup#1\@@endlink}%
\providecommand \@sanitize@url [0]{\catcode `\\12\catcode `\$12\catcode
  `\&12\catcode `\#12\catcode `\^12\catcode `\_12\catcode `\%12\relax}%
\providecommand \@@startlink[1]{}%
\providecommand \@@endlink[0]{}%
\providecommand \url  [0]{\begingroup\@sanitize@url \@url }%
\providecommand \@url [1]{\endgroup\@href {#1}{\urlprefix }}%
\providecommand \urlprefix  [0]{URL }%
\providecommand \Eprint [0]{\href }%
\providecommand \doibase [0]{https://doi.org/}%
\providecommand \selectlanguage [0]{\@gobble}%
\providecommand \bibinfo  [0]{\@secondoftwo}%
\providecommand \bibfield  [0]{\@secondoftwo}%
\providecommand \translation [1]{[#1]}%
\providecommand \BibitemOpen [0]{}%
\providecommand \bibitemStop [0]{}%
\providecommand \bibitemNoStop [0]{.\EOS\space}%
\providecommand \EOS [0]{\spacefactor3000\relax}%
\providecommand \BibitemShut  [1]{\csname bibitem#1\endcsname}%
\let\auto@bib@innerbib\@empty
\bibitem [{\citenamefont {Gulian}\ \emph {et~al.}(2018)\citenamefont {Gulian},
  \citenamefont {Nikoghosyan}, \citenamefont {Gulian},\ and\ \citenamefont
  {Melkonyan}}]{Gulian18}%
  \BibitemOpen
  \bibfield  {author} {\bibinfo {author} {\bibfnamefont {A.~M.}\ \bibnamefont
  {Gulian}}, \bibinfo {author} {\bibfnamefont {V.~R.}\ \bibnamefont
  {Nikoghosyan}}, \bibinfo {author} {\bibfnamefont {E.~D.}\ \bibnamefont
  {Gulian}},\ and\ \bibinfo {author} {\bibfnamefont {G.~G.}\ \bibnamefont
  {Melkonyan}},\ }\bibfield  {title} {\bibinfo {title} {Quasi-local action of
  curl-less vector potential on vortex dynamics in superconductors},\
  }\href@noop {} {\bibfield  {journal} {\bibinfo  {journal} {Physics Letters
  A}\ }\textbf {\bibinfo {volume} {382}},\ \bibinfo {pages} {1058} (\bibinfo
  {year} {2018})}\BibitemShut {NoStop}%
\bibitem [{\citenamefont {Tokura}\ and\ \citenamefont
  {Nagaosa}(2018)}]{Tokura18}%
  \BibitemOpen
  \bibfield  {author} {\bibinfo {author} {\bibfnamefont {Y.}~\bibnamefont
  {Tokura}}\ and\ \bibinfo {author} {\bibfnamefont {N.}~\bibnamefont
  {Nagaosa}},\ }\bibfield  {title} {\bibinfo {title} {Nonreciprocal responses
  from non-centrosymmetric quantum materials},\ }\href@noop {} {\bibfield
  {journal} {\bibinfo  {journal} {Nature Communications}\ }\textbf {\bibinfo
  {volume} {9}},\ \bibinfo {pages} {3740} (\bibinfo {year} {2018})}\BibitemShut
  {NoStop}%
\bibitem [{\citenamefont {Wakatsuki}\ and\ \citenamefont
  {Nagaosa}(2018)}]{Wakatsuki18}%
  \BibitemOpen
  \bibfield  {author} {\bibinfo {author} {\bibfnamefont {R.}~\bibnamefont
  {Wakatsuki}}\ and\ \bibinfo {author} {\bibfnamefont {N.}~\bibnamefont
  {Nagaosa}},\ }\bibfield  {title} {\bibinfo {title} {{Nonreciprocal Current in
  Noncentrosymmetric Rashba Superconductors}},\ }\href@noop {} {\bibfield
  {journal} {\bibinfo  {journal} {Phys. Rev. Lett.}\ }\textbf {\bibinfo
  {volume} {121}},\ \bibinfo {pages} {026601} (\bibinfo {year}
  {2018})}\BibitemShut {NoStop}%
\bibitem [{\citenamefont {Hoshino}\ \emph {et~al.}(2018)\citenamefont
  {Hoshino}, \citenamefont {Wakatsuki}, \citenamefont {Hamamoto},\ and\
  \citenamefont {Nagaosa}}]{Hoshino18}%
  \BibitemOpen
  \bibfield  {author} {\bibinfo {author} {\bibfnamefont {S.}~\bibnamefont
  {Hoshino}}, \bibinfo {author} {\bibfnamefont {R.}~\bibnamefont {Wakatsuki}},
  \bibinfo {author} {\bibfnamefont {K.}~\bibnamefont {Hamamoto}},\ and\
  \bibinfo {author} {\bibfnamefont {N.}~\bibnamefont {Nagaosa}},\ }\bibfield
  {title} {\bibinfo {title} {Nonreciprocal charge transport in two-dimensional
  noncentrosymmetric superconductors},\ }\href@noop {} {\bibfield  {journal}
  {\bibinfo  {journal} {Phys. Rev. B}\ }\textbf {\bibinfo {volume} {98}},\
  \bibinfo {pages} {054510} (\bibinfo {year} {2018})}\BibitemShut {NoStop}%
\bibitem [{\citenamefont {Ando}\ \emph {et~al.}(2020)\citenamefont {Ando},
  \citenamefont {Miyasaka}, \citenamefont {Li}, \citenamefont {Ishizuka},
  \citenamefont {Arakawa}, \citenamefont {Shiota}, \citenamefont {Moriyama},
  \citenamefont {Yanase},\ and\ \citenamefont {Ono}}]{Ando20}%
  \BibitemOpen
  \bibfield  {author} {\bibinfo {author} {\bibfnamefont {F.}~\bibnamefont
  {Ando}}, \bibinfo {author} {\bibfnamefont {Y.}~\bibnamefont {Miyasaka}},
  \bibinfo {author} {\bibfnamefont {T.}~\bibnamefont {Li}}, \bibinfo {author}
  {\bibfnamefont {J.}~\bibnamefont {Ishizuka}}, \bibinfo {author}
  {\bibfnamefont {T.}~\bibnamefont {Arakawa}}, \bibinfo {author} {\bibfnamefont
  {Y.}~\bibnamefont {Shiota}}, \bibinfo {author} {\bibfnamefont
  {T.}~\bibnamefont {Moriyama}}, \bibinfo {author} {\bibfnamefont
  {Y.}~\bibnamefont {Yanase}},\ and\ \bibinfo {author} {\bibfnamefont
  {T.}~\bibnamefont {Ono}},\ }\bibfield  {title} {\bibinfo {title} {Observation
  of superconducting diode effect},\ }\href@noop {} {\bibfield  {journal}
  {\bibinfo  {journal} {Nature}\ }\textbf {\bibinfo {volume} {584}},\ \bibinfo
  {pages} {373} (\bibinfo {year} {2020})}\BibitemShut {NoStop}%
\bibitem [{\citenamefont {Ideue}\ and\ \citenamefont {Iwasa}(2020)}]{Ideue20}%
  \BibitemOpen
  \bibfield  {author} {\bibinfo {author} {\bibfnamefont {T.}~\bibnamefont
  {Ideue}}\ and\ \bibinfo {author} {\bibfnamefont {Y.}~\bibnamefont {Iwasa}},\
  }\bibfield  {title} {\bibinfo {title} {One-way supercurrent achieved in an
  electrically polar film},\ }\href@noop {} {\bibfield  {journal} {\bibinfo
  {journal} {Nature}\ }\textbf {\bibinfo {volume} {584}},\ \bibinfo {pages}
  {349} (\bibinfo {year} {2020})}\BibitemShut {NoStop}%
\bibitem [{\citenamefont {Baumgartner}\ \emph
  {et~al.}(2022{\natexlab{a}})\citenamefont {Baumgartner}, \citenamefont
  {Fuchs}, \citenamefont {Costa}, \citenamefont {Reinhardt}, \citenamefont
  {Gronin}, \citenamefont {Gardner}, \citenamefont {Lindemann}, \citenamefont
  {Manfra}, \citenamefont {Faria~Junior}, \citenamefont {Kochan}, \citenamefont
  {Fabian}, \citenamefont {Paradiso},\ and\ \citenamefont
  {Strunk}}]{BaumgartnerNN}%
  \BibitemOpen
  \bibfield  {author} {\bibinfo {author} {\bibfnamefont {C.}~\bibnamefont
  {Baumgartner}}, \bibinfo {author} {\bibfnamefont {L.}~\bibnamefont {Fuchs}},
  \bibinfo {author} {\bibfnamefont {A.}~\bibnamefont {Costa}}, \bibinfo
  {author} {\bibfnamefont {S.}~\bibnamefont {Reinhardt}}, \bibinfo {author}
  {\bibfnamefont {S.}~\bibnamefont {Gronin}}, \bibinfo {author} {\bibfnamefont
  {G.~C.}\ \bibnamefont {Gardner}}, \bibinfo {author} {\bibfnamefont
  {T.}~\bibnamefont {Lindemann}}, \bibinfo {author} {\bibfnamefont {M.~J.}\
  \bibnamefont {Manfra}}, \bibinfo {author} {\bibfnamefont {P.~E.}\
  \bibnamefont {Faria~Junior}}, \bibinfo {author} {\bibfnamefont
  {D.}~\bibnamefont {Kochan}}, \bibinfo {author} {\bibfnamefont
  {J.}~\bibnamefont {Fabian}}, \bibinfo {author} {\bibfnamefont
  {N.}~\bibnamefont {Paradiso}},\ and\ \bibinfo {author} {\bibfnamefont
  {C.}~\bibnamefont {Strunk}},\ }\bibfield  {title} {\bibinfo {title}
  {{Supercurrent rectification and magnetochiral effects in symmetric Josephson
  junctions}},\ }\href@noop {} {\bibfield  {journal} {\bibinfo  {journal}
  {Nature Nanotechnology}\ }\textbf {\bibinfo {volume} {17}},\ \bibinfo {pages}
  {39} (\bibinfo {year} {2022}{\natexlab{a}})}\BibitemShut {NoStop}%
\bibitem [{\citenamefont {Wu}\ \emph {et~al.}(2022)\citenamefont {Wu},
  \citenamefont {Wang}, \citenamefont {Xu}, \citenamefont {Sivakumar},
  \citenamefont {Pasco}, \citenamefont {Filippozzi}, \citenamefont {Parkin},
  \citenamefont {Zeng}, \citenamefont {McQueen},\ and\ \citenamefont
  {Ali}}]{Wu22}%
  \BibitemOpen
  \bibfield  {author} {\bibinfo {author} {\bibfnamefont {H.}~\bibnamefont
  {Wu}}, \bibinfo {author} {\bibfnamefont {Y.}~\bibnamefont {Wang}}, \bibinfo
  {author} {\bibfnamefont {Y.}~\bibnamefont {Xu}}, \bibinfo {author}
  {\bibfnamefont {P.~K.}\ \bibnamefont {Sivakumar}}, \bibinfo {author}
  {\bibfnamefont {C.}~\bibnamefont {Pasco}}, \bibinfo {author} {\bibfnamefont
  {U.}~\bibnamefont {Filippozzi}}, \bibinfo {author} {\bibfnamefont {S.~S.~P.}\
  \bibnamefont {Parkin}}, \bibinfo {author} {\bibfnamefont {Y.-J.}\
  \bibnamefont {Zeng}}, \bibinfo {author} {\bibfnamefont {T.}~\bibnamefont
  {McQueen}},\ and\ \bibinfo {author} {\bibfnamefont {M.~N.}\ \bibnamefont
  {Ali}},\ }\bibfield  {title} {\bibinfo {title} {{The field-free Josephson
  diode in a van der Waals heterostructure}},\ }\href@noop {} {\bibfield
  {journal} {\bibinfo  {journal} {Nature}\ }\textbf {\bibinfo {volume} {604}},\
  \bibinfo {pages} {653} (\bibinfo {year} {2022})}\BibitemShut {NoStop}%
\bibitem [{\citenamefont {Strambini}\ \emph {et~al.}(2022)\citenamefont
  {Strambini}, \citenamefont {Spies}, \citenamefont {Ligato}, \citenamefont
  {Ili{\'{c}}}, \citenamefont {Rouco}, \citenamefont {Gonz{\'a}lez-Orellana},
  \citenamefont {Ilyn}, \citenamefont {Rogero}, \citenamefont {Bergeret},
  \citenamefont {Moodera}, \citenamefont {Virtanen}, \citenamefont
  {Heikkil{\"a}},\ and\ \citenamefont {Giazotto}}]{Strambini22}%
  \BibitemOpen
  \bibfield  {author} {\bibinfo {author} {\bibfnamefont {E.}~\bibnamefont
  {Strambini}}, \bibinfo {author} {\bibfnamefont {M.}~\bibnamefont {Spies}},
  \bibinfo {author} {\bibfnamefont {N.}~\bibnamefont {Ligato}}, \bibinfo
  {author} {\bibfnamefont {S.}~\bibnamefont {Ili{\'{c}}}}, \bibinfo {author}
  {\bibfnamefont {M.}~\bibnamefont {Rouco}}, \bibinfo {author} {\bibfnamefont
  {C.}~\bibnamefont {Gonz{\'a}lez-Orellana}}, \bibinfo {author} {\bibfnamefont
  {M.}~\bibnamefont {Ilyn}}, \bibinfo {author} {\bibfnamefont {C.}~\bibnamefont
  {Rogero}}, \bibinfo {author} {\bibfnamefont {F.~S.}\ \bibnamefont
  {Bergeret}}, \bibinfo {author} {\bibfnamefont {J.~S.}\ \bibnamefont
  {Moodera}}, \bibinfo {author} {\bibfnamefont {P.}~\bibnamefont {Virtanen}},
  \bibinfo {author} {\bibfnamefont {T.~T.}\ \bibnamefont {Heikkil{\"a}}},\ and\
  \bibinfo {author} {\bibfnamefont {F.}~\bibnamefont {Giazotto}},\ }\bibfield
  {title} {\bibinfo {title} {Superconducting spintronic tunnel diode},\
  }\href@noop {} {\bibfield  {journal} {\bibinfo  {journal} {Nature
  Communications}\ }\textbf {\bibinfo {volume} {13}},\ \bibinfo {pages} {2431}
  (\bibinfo {year} {2022})}\BibitemShut {NoStop}%
\bibitem [{\citenamefont {Wakatsuki}\ \emph {et~al.}(2017)\citenamefont
  {Wakatsuki}, \citenamefont {Saito}, \citenamefont {Hoshino}, \citenamefont
  {Itahashi}, \citenamefont {Ideue}, \citenamefont {Ezawa}, \citenamefont
  {Iwasa},\ and\ \citenamefont {Nagaosa}}]{Wakatsuki17}%
  \BibitemOpen
  \bibfield  {author} {\bibinfo {author} {\bibfnamefont {R.}~\bibnamefont
  {Wakatsuki}}, \bibinfo {author} {\bibfnamefont {Y.}~\bibnamefont {Saito}},
  \bibinfo {author} {\bibfnamefont {S.}~\bibnamefont {Hoshino}}, \bibinfo
  {author} {\bibfnamefont {Y.~M.}\ \bibnamefont {Itahashi}}, \bibinfo {author}
  {\bibfnamefont {T.}~\bibnamefont {Ideue}}, \bibinfo {author} {\bibfnamefont
  {M.}~\bibnamefont {Ezawa}}, \bibinfo {author} {\bibfnamefont
  {Y.}~\bibnamefont {Iwasa}},\ and\ \bibinfo {author} {\bibfnamefont
  {N.}~\bibnamefont {Nagaosa}},\ }\bibfield  {title} {\bibinfo {title}
  {Nonreciprocal charge transport in noncentrosymmetric superconductors},\
  }\href@noop {} {\bibfield  {journal} {\bibinfo  {journal} {Science Advances}\
  }\textbf {\bibinfo {volume} {3}},\ \bibinfo {pages} {e1602390} (\bibinfo
  {year} {2017})}\BibitemShut {NoStop}%
\bibitem [{\citenamefont {Shin}\ \emph {et~al.}(2021)\citenamefont {Shin},
  \citenamefont {Son}, \citenamefont {Yun}, \citenamefont {Park}, \citenamefont
  {Zhang}, \citenamefont {Shin}, \citenamefont {Park},\ and\ \citenamefont
  {Kim}}]{Shin21}%
  \BibitemOpen
  \bibfield  {author} {\bibinfo {author} {\bibfnamefont {J.}~\bibnamefont
  {Shin}}, \bibinfo {author} {\bibfnamefont {S.}~\bibnamefont {Son}}, \bibinfo
  {author} {\bibfnamefont {J.}~\bibnamefont {Yun}}, \bibinfo {author}
  {\bibfnamefont {G.}~\bibnamefont {Park}}, \bibinfo {author} {\bibfnamefont
  {K.}~\bibnamefont {Zhang}}, \bibinfo {author} {\bibfnamefont {Y.~J.}\
  \bibnamefont {Shin}}, \bibinfo {author} {\bibfnamefont {J.-G.}\ \bibnamefont
  {Park}},\ and\ \bibinfo {author} {\bibfnamefont {D.}~\bibnamefont {Kim}},\
  }\bibfield  {title} {\bibinfo {title} {{Magnetic Proximity-Induced
  Superconducting Diode Effect and Infinite Magnetoresistance in van der Waals
  Heterostructure}},\ }\href@noop {} {\bibfield  {journal} {\bibinfo  {journal}
  {arXiv.2111.05627}\ } (\bibinfo {year} {2021})}\BibitemShut {NoStop}%
\bibitem [{\citenamefont {Baumgartner}\ \emph
  {et~al.}(2022{\natexlab{b}})\citenamefont {Baumgartner}, \citenamefont
  {Fuchs}, \citenamefont {Costa}, \citenamefont {Pic{\'{o}}-Cort{\'{e}}s},
  \citenamefont {Reinhardt}, \citenamefont {Gronin}, \citenamefont {Gardner},
  \citenamefont {Lindemann}, \citenamefont {Manfra}, \citenamefont {Junior},
  \citenamefont {Kochan}, \citenamefont {Fabian}, \citenamefont {Paradiso},\
  and\ \citenamefont {Strunk}}]{BaumgartnerIOP}%
  \BibitemOpen
  \bibfield  {author} {\bibinfo {author} {\bibfnamefont {C.}~\bibnamefont
  {Baumgartner}}, \bibinfo {author} {\bibfnamefont {L.}~\bibnamefont {Fuchs}},
  \bibinfo {author} {\bibfnamefont {A.}~\bibnamefont {Costa}}, \bibinfo
  {author} {\bibfnamefont {J.}~\bibnamefont {Pic{\'{o}}-Cort{\'{e}}s}},
  \bibinfo {author} {\bibfnamefont {S.}~\bibnamefont {Reinhardt}}, \bibinfo
  {author} {\bibfnamefont {S.}~\bibnamefont {Gronin}}, \bibinfo {author}
  {\bibfnamefont {G.~C.}\ \bibnamefont {Gardner}}, \bibinfo {author}
  {\bibfnamefont {T.}~\bibnamefont {Lindemann}}, \bibinfo {author}
  {\bibfnamefont {M.~J.}\ \bibnamefont {Manfra}}, \bibinfo {author}
  {\bibfnamefont {P.~E.~F.}\ \bibnamefont {Junior}}, \bibinfo {author}
  {\bibfnamefont {D.}~\bibnamefont {Kochan}}, \bibinfo {author} {\bibfnamefont
  {J.}~\bibnamefont {Fabian}}, \bibinfo {author} {\bibfnamefont
  {N.}~\bibnamefont {Paradiso}},\ and\ \bibinfo {author} {\bibfnamefont
  {C.}~\bibnamefont {Strunk}},\ }\bibfield  {title} {\bibinfo {title} {{Effect
  of Rashba and Dresselhaus spin{\textendash}orbit coupling on supercurrent
  rectification and magnetochiral anisotropy of ballistic Josephson
  junctions}},\ }\href@noop {} {\bibfield  {journal} {\bibinfo  {journal}
  {Journal of Physics: Condensed Matter}\ }\textbf {\bibinfo {volume} {34}},\
  \bibinfo {pages} {154005} (\bibinfo {year} {2022}{\natexlab{b}})}\BibitemShut
  {NoStop}%
\bibitem [{\citenamefont {Bauriedl}\ \emph {et~al.}(2022)\citenamefont
  {Bauriedl}, \citenamefont {B{\"a}uml}, \citenamefont {Fuchs}, \citenamefont
  {Baumgartner}, \citenamefont {Paulik}, \citenamefont {Bauer}, \citenamefont
  {Lin}, \citenamefont {Lupton}, \citenamefont {Taniguchi}, \citenamefont
  {Watanabe}, \citenamefont {Strunk},\ and\ \citenamefont
  {Paradiso}}]{Bauriedl22}%
  \BibitemOpen
  \bibfield  {author} {\bibinfo {author} {\bibfnamefont {L.}~\bibnamefont
  {Bauriedl}}, \bibinfo {author} {\bibfnamefont {C.}~\bibnamefont {B{\"a}uml}},
  \bibinfo {author} {\bibfnamefont {L.}~\bibnamefont {Fuchs}}, \bibinfo
  {author} {\bibfnamefont {C.}~\bibnamefont {Baumgartner}}, \bibinfo {author}
  {\bibfnamefont {N.}~\bibnamefont {Paulik}}, \bibinfo {author} {\bibfnamefont
  {J.~M.}\ \bibnamefont {Bauer}}, \bibinfo {author} {\bibfnamefont {K.-Q.}\
  \bibnamefont {Lin}}, \bibinfo {author} {\bibfnamefont {J.~M.}\ \bibnamefont
  {Lupton}}, \bibinfo {author} {\bibfnamefont {T.}~\bibnamefont {Taniguchi}},
  \bibinfo {author} {\bibfnamefont {K.}~\bibnamefont {Watanabe}}, \bibinfo
  {author} {\bibfnamefont {C.}~\bibnamefont {Strunk}},\ and\ \bibinfo {author}
  {\bibfnamefont {N.}~\bibnamefont {Paradiso}},\ }\bibfield  {title} {\bibinfo
  {title} {{Supercurrent diode effect and magnetochiral anisotropy in few-layer
  NbSe$_2$}},\ }\href@noop {} {\bibfield  {journal} {\bibinfo  {journal}
  {Nature Communications}\ }\textbf {\bibinfo {volume} {13}},\ \bibinfo {pages}
  {4266} (\bibinfo {year} {2022})}\BibitemShut {NoStop}%
\bibitem [{\citenamefont {Yuan}\ and\ \citenamefont {Fu}(2022)}]{Yuan22}%
  \BibitemOpen
  \bibfield  {author} {\bibinfo {author} {\bibfnamefont {N.~F.~Q.}\
  \bibnamefont {Yuan}}\ and\ \bibinfo {author} {\bibfnamefont {L.}~\bibnamefont
  {Fu}},\ }\bibfield  {title} {\bibinfo {title} {Supercurrent diode effect and
  finite-momentum superconductors},\ }\href@noop {} {\bibfield  {journal}
  {\bibinfo  {journal} {Proceedings of the National Academy of Sciences}\
  }\textbf {\bibinfo {volume} {119}},\ \bibinfo {pages} {e2119548119} (\bibinfo
  {year} {2022})}\BibitemShut {NoStop}%
\bibitem [{\citenamefont {He}\ \emph {et~al.}(2022)\citenamefont {He},
  \citenamefont {Tanaka},\ and\ \citenamefont {Nagaosa}}]{He22}%
  \BibitemOpen
  \bibfield  {author} {\bibinfo {author} {\bibfnamefont {J.~J.}\ \bibnamefont
  {He}}, \bibinfo {author} {\bibfnamefont {Y.}~\bibnamefont {Tanaka}},\ and\
  \bibinfo {author} {\bibfnamefont {N.}~\bibnamefont {Nagaosa}},\ }\bibfield
  {title} {\bibinfo {title} {A phenomenological theory of superconductor
  diodes},\ }\href@noop {} {\bibfield  {journal} {\bibinfo  {journal} {New
  Journal of Physics}\ }\textbf {\bibinfo {volume} {24}},\ \bibinfo {pages}
  {053014} (\bibinfo {year} {2022})}\BibitemShut {NoStop}%
\bibitem [{\citenamefont {Lin}\ \emph {et~al.}(2022)\citenamefont {Lin},
  \citenamefont {Siriviboon}, \citenamefont {Scammell}, \citenamefont {Liu},
  \citenamefont {Rhodes}, \citenamefont {Watanabe}, \citenamefont {Taniguchi},
  \citenamefont {Hone}, \citenamefont {Scheurer},\ and\ \citenamefont
  {Li}}]{Lin22}%
  \BibitemOpen
  \bibfield  {author} {\bibinfo {author} {\bibfnamefont {J.-X.}\ \bibnamefont
  {Lin}}, \bibinfo {author} {\bibfnamefont {P.}~\bibnamefont {Siriviboon}},
  \bibinfo {author} {\bibfnamefont {H.~D.}\ \bibnamefont {Scammell}}, \bibinfo
  {author} {\bibfnamefont {S.}~\bibnamefont {Liu}}, \bibinfo {author}
  {\bibfnamefont {D.}~\bibnamefont {Rhodes}}, \bibinfo {author} {\bibfnamefont
  {K.}~\bibnamefont {Watanabe}}, \bibinfo {author} {\bibfnamefont
  {T.}~\bibnamefont {Taniguchi}}, \bibinfo {author} {\bibfnamefont
  {J.}~\bibnamefont {Hone}}, \bibinfo {author} {\bibfnamefont {M.~S.}\
  \bibnamefont {Scheurer}},\ and\ \bibinfo {author} {\bibfnamefont {J.~I.~A.}\
  \bibnamefont {Li}},\ }\bibfield  {title} {\bibinfo {title} {Zero-field
  superconducting diode effect in small-twist-angle trilayer graphene},\
  }\href@noop {} {\bibfield  {journal} {\bibinfo  {journal} {Nature Physics}\ }
  (\bibinfo {year} {2022})}\BibitemShut {NoStop}%
\bibitem [{\citenamefont {Ili\ifmmode~\acute{c}\else \'{c}\fi{}}\ and\
  \citenamefont {Bergeret}(2022)}]{Ilic22}%
  \BibitemOpen
  \bibfield  {author} {\bibinfo {author} {\bibfnamefont {S.}~\bibnamefont
  {Ili\ifmmode~\acute{c}\else \'{c}\fi{}}}\ and\ \bibinfo {author}
  {\bibfnamefont {F.~S.}\ \bibnamefont {Bergeret}},\ }\bibfield  {title}
  {\bibinfo {title} {{Theory of the Supercurrent Diode Effect in Rashba
  Superconductors with Arbitrary Disorder}},\ }\href@noop {} {\bibfield
  {journal} {\bibinfo  {journal} {Phys. Rev. Lett.}\ }\textbf {\bibinfo
  {volume} {128}},\ \bibinfo {pages} {177001} (\bibinfo {year}
  {2022})}\BibitemShut {NoStop}%
\bibitem [{\citenamefont {Karabassov}\ \emph {et~al.}(2022)\citenamefont
  {Karabassov}, \citenamefont {Bobkova}, \citenamefont {Golubov},\ and\
  \citenamefont {Vasenko}}]{Karabassov22}%
  \BibitemOpen
  \bibfield  {author} {\bibinfo {author} {\bibfnamefont {T.}~\bibnamefont
  {Karabassov}}, \bibinfo {author} {\bibfnamefont {I.~V.}\ \bibnamefont
  {Bobkova}}, \bibinfo {author} {\bibfnamefont {A.~A.}\ \bibnamefont
  {Golubov}},\ and\ \bibinfo {author} {\bibfnamefont {A.~S.}\ \bibnamefont
  {Vasenko}},\ }\bibfield  {title} {\bibinfo {title} {{Hybrid helical state and
  superconducting diode effect in S/F/TI heterostructures}},\ }\href@noop {}
  {\bibfield  {journal} {\bibinfo  {journal} {arXiv.2203.15608}\ } (\bibinfo
  {year} {2022})}\BibitemShut {NoStop}%
\bibitem [{\citenamefont {Daido}\ \emph {et~al.}(2022)\citenamefont {Daido},
  \citenamefont {Ikeda},\ and\ \citenamefont {Yanase}}]{Daido22}%
  \BibitemOpen
  \bibfield  {author} {\bibinfo {author} {\bibfnamefont {A.}~\bibnamefont
  {Daido}}, \bibinfo {author} {\bibfnamefont {Y.}~\bibnamefont {Ikeda}},\ and\
  \bibinfo {author} {\bibfnamefont {Y.}~\bibnamefont {Yanase}},\ }\bibfield
  {title} {\bibinfo {title} {{Intrinsic Superconducting Diode Effect}},\
  }\href@noop {} {\bibfield  {journal} {\bibinfo  {journal} {Phys. Rev. Lett.}\
  }\textbf {\bibinfo {volume} {128}},\ \bibinfo {pages} {037001} (\bibinfo
  {year} {2022})}\BibitemShut {NoStop}%
\bibitem [{\citenamefont {Anwar}\ \emph {et~al.}(2022)\citenamefont {Anwar},
  \citenamefont {Nakamura}, \citenamefont {Ishiguro}, \citenamefont {Arif},
  \citenamefont {Robinson}, \citenamefont {Yonezawa}, \citenamefont {Sigrist},\
  and\ \citenamefont {Maeno}}]{Anwar22}%
  \BibitemOpen
  \bibfield  {author} {\bibinfo {author} {\bibfnamefont {M.~S.}\ \bibnamefont
  {Anwar}}, \bibinfo {author} {\bibfnamefont {T.}~\bibnamefont {Nakamura}},
  \bibinfo {author} {\bibfnamefont {R.}~\bibnamefont {Ishiguro}}, \bibinfo
  {author} {\bibfnamefont {S.}~\bibnamefont {Arif}}, \bibinfo {author}
  {\bibfnamefont {J.~W.~A.}\ \bibnamefont {Robinson}}, \bibinfo {author}
  {\bibfnamefont {S.}~\bibnamefont {Yonezawa}}, \bibinfo {author}
  {\bibfnamefont {M.}~\bibnamefont {Sigrist}},\ and\ \bibinfo {author}
  {\bibfnamefont {Y.}~\bibnamefont {Maeno}},\ }\bibfield  {title} {\bibinfo
  {title} {{Spontaneous superconducting diode effect in non-magnetic
  Nb/Ru/Sr$_2$RuO$_4$ topological junctions}},\ }\href@noop {} {\bibfield
  {journal} {\bibinfo  {journal} {arXiv.2211.14626}\ } (\bibinfo {year}
  {2022})}\BibitemShut {NoStop}%
\bibitem [{\citenamefont {Souto}\ \emph {et~al.}(2022)\citenamefont {Souto},
  \citenamefont {Leijnse},\ and\ \citenamefont {Schrade}}]{Souto22}%
  \BibitemOpen
  \bibfield  {author} {\bibinfo {author} {\bibfnamefont {R.~S.}\ \bibnamefont
  {Souto}}, \bibinfo {author} {\bibfnamefont {M.}~\bibnamefont {Leijnse}},\
  and\ \bibinfo {author} {\bibfnamefont {C.}~\bibnamefont {Schrade}},\
  }\bibfield  {title} {\bibinfo {title} {{The Josephson diode effect in
  supercurrent interferometers}},\ }\href@noop {} {\bibfield  {journal}
  {\bibinfo  {journal} {arXiv.2205.04469}\ } (\bibinfo {year}
  {2022})}\BibitemShut {NoStop}%
\bibitem [{\citenamefont {Halterman}\ \emph {et~al.}(2022)\citenamefont
  {Halterman}, \citenamefont {Alidoust}, \citenamefont {Smith},\ and\
  \citenamefont {Starr}}]{Halterman22}%
  \BibitemOpen
  \bibfield  {author} {\bibinfo {author} {\bibfnamefont {K.}~\bibnamefont
  {Halterman}}, \bibinfo {author} {\bibfnamefont {M.}~\bibnamefont {Alidoust}},
  \bibinfo {author} {\bibfnamefont {R.}~\bibnamefont {Smith}},\ and\ \bibinfo
  {author} {\bibfnamefont {S.}~\bibnamefont {Starr}},\ }\bibfield  {title}
  {\bibinfo {title} {Supercurrent diode effect, spin torques, and robust
  zero-energy peak in planar half-metallic trilayers},\ }\href@noop {}
  {\bibfield  {journal} {\bibinfo  {journal} {Phys. Rev. B}\ }\textbf {\bibinfo
  {volume} {105}},\ \bibinfo {pages} {104508} (\bibinfo {year}
  {2022})}\BibitemShut {NoStop}%
\bibitem [{\citenamefont {Hou}\ \emph {et~al.}(2022)\citenamefont {Hou},
  \citenamefont {Nichele}, \citenamefont {Chi}, \citenamefont {Lodesani},
  \citenamefont {Wu}, \citenamefont {Ritter}, \citenamefont {Haxell},
  \citenamefont {Davydova}, \citenamefont {Ilić}, \citenamefont {Bergeret},
  \citenamefont {Kamra}, \citenamefont {Fu}, \citenamefont {Lee},\ and\
  \citenamefont {Moodera}}]{Hou22}%
  \BibitemOpen
  \bibfield  {author} {\bibinfo {author} {\bibfnamefont {Y.}~\bibnamefont
  {Hou}}, \bibinfo {author} {\bibfnamefont {F.}~\bibnamefont {Nichele}},
  \bibinfo {author} {\bibfnamefont {H.}~\bibnamefont {Chi}}, \bibinfo {author}
  {\bibfnamefont {A.}~\bibnamefont {Lodesani}}, \bibinfo {author}
  {\bibfnamefont {Y.}~\bibnamefont {Wu}}, \bibinfo {author} {\bibfnamefont
  {M.~F.}\ \bibnamefont {Ritter}}, \bibinfo {author} {\bibfnamefont {D.~Z.}\
  \bibnamefont {Haxell}}, \bibinfo {author} {\bibfnamefont {M.}~\bibnamefont
  {Davydova}}, \bibinfo {author} {\bibfnamefont {S.}~\bibnamefont {Ilić}},
  \bibinfo {author} {\bibfnamefont {F.~S.}\ \bibnamefont {Bergeret}}, \bibinfo
  {author} {\bibfnamefont {A.}~\bibnamefont {Kamra}}, \bibinfo {author}
  {\bibfnamefont {L.}~\bibnamefont {Fu}}, \bibinfo {author} {\bibfnamefont
  {P.~A.}\ \bibnamefont {Lee}},\ and\ \bibinfo {author} {\bibfnamefont {J.~S.}\
  \bibnamefont {Moodera}},\ }\bibfield  {title} {\bibinfo {title} {{Ubiquitous
  Superconducting Diode Effect in Superconductor Thin Films}},\ }\href@noop {}
  {\bibfield  {journal} {\bibinfo  {journal} {arXiv.2205.09276}\ } (\bibinfo
  {year} {2022})}\BibitemShut {NoStop}%
\bibitem [{\citenamefont {Suri}\ \emph {et~al.}(2022)\citenamefont {Suri},
  \citenamefont {Kamra}, \citenamefont {Meier}, \citenamefont {Kronseder},
  \citenamefont {Belzig}, \citenamefont {Back},\ and\ \citenamefont
  {Strunk}}]{Suri22}%
  \BibitemOpen
  \bibfield  {author} {\bibinfo {author} {\bibfnamefont {D.}~\bibnamefont
  {Suri}}, \bibinfo {author} {\bibfnamefont {A.}~\bibnamefont {Kamra}},
  \bibinfo {author} {\bibfnamefont {T.~N.~G.}\ \bibnamefont {Meier}}, \bibinfo
  {author} {\bibfnamefont {M.}~\bibnamefont {Kronseder}}, \bibinfo {author}
  {\bibfnamefont {W.}~\bibnamefont {Belzig}}, \bibinfo {author} {\bibfnamefont
  {C.~H.}\ \bibnamefont {Back}},\ and\ \bibinfo {author} {\bibfnamefont
  {C.}~\bibnamefont {Strunk}},\ }\bibfield  {title} {\bibinfo {title}
  {Non-reciprocity of vortex-limited critical current in conventional
  superconducting micro-bridges},\ }\href@noop {} {\bibfield  {journal}
  {\bibinfo  {journal} {Applied Physics Letters}\ }\textbf {\bibinfo {volume}
  {121}},\ \bibinfo {pages} {102601} (\bibinfo {year} {2022})}\BibitemShut
  {NoStop}%
\bibitem [{\citenamefont {Hope}\ \emph {et~al.}(2021)\citenamefont {Hope},
  \citenamefont {Amundsen}, \citenamefont {Suri}, \citenamefont {Moodera},\
  and\ \citenamefont {Kamra}}]{Hope21}%
  \BibitemOpen
  \bibfield  {author} {\bibinfo {author} {\bibfnamefont {M.~K.}\ \bibnamefont
  {Hope}}, \bibinfo {author} {\bibfnamefont {M.}~\bibnamefont {Amundsen}},
  \bibinfo {author} {\bibfnamefont {D.}~\bibnamefont {Suri}}, \bibinfo {author}
  {\bibfnamefont {J.~S.}\ \bibnamefont {Moodera}},\ and\ \bibinfo {author}
  {\bibfnamefont {A.}~\bibnamefont {Kamra}},\ }\bibfield  {title} {\bibinfo
  {title} {{Interfacial control of vortex-limited critical current in type-II
  superconductor films}},\ }\href@noop {} {\bibfield  {journal} {\bibinfo
  {journal} {Phys. Rev. B}\ }\textbf {\bibinfo {volume} {104}},\ \bibinfo
  {pages} {184512} (\bibinfo {year} {2021})}\BibitemShut {NoStop}%
\bibitem [{\citenamefont {Vodolazov}\ and\ \citenamefont
  {Peeters}(2005)}]{Vodolazov05}%
  \BibitemOpen
  \bibfield  {author} {\bibinfo {author} {\bibfnamefont {D.~Y.}\ \bibnamefont
  {Vodolazov}}\ and\ \bibinfo {author} {\bibfnamefont {F.~M.}\ \bibnamefont
  {Peeters}},\ }\bibfield  {title} {\bibinfo {title} {Superconducting rectifier
  based on the asymmetric surface barrier effect},\ }\href@noop {} {\bibfield
  {journal} {\bibinfo  {journal} {Phys. Rev. B}\ }\textbf {\bibinfo {volume}
  {72}},\ \bibinfo {pages} {172508} (\bibinfo {year} {2005})}\BibitemShut
  {NoStop}%
\bibitem [{\citenamefont {Shmidt}(1970{\natexlab{a}})}]{Shmidt1}%
  \BibitemOpen
  \bibfield  {author} {\bibinfo {author} {\bibfnamefont {V.~V.}\ \bibnamefont
  {Shmidt}},\ }\bibfield  {title} {\bibinfo {title} {The critical current in
  superconducting films},\ }\href@noop {} {\bibfield  {journal} {\bibinfo
  {journal} {Sov. Phys. JETP}\ }\textbf {\bibinfo {volume} {30}},\ \bibinfo
  {pages} {1137} (\bibinfo {year} {1970}{\natexlab{a}})}\BibitemShut {NoStop}%
\bibitem [{\citenamefont {Shmidt}(1970{\natexlab{b}})}]{Shmidt2}%
  \BibitemOpen
  \bibfield  {author} {\bibinfo {author} {\bibfnamefont {V.~V.}\ \bibnamefont
  {Shmidt}},\ }\bibfield  {title} {\bibinfo {title} {Critical currents in
  superconductors},\ }\href@noop {} {\bibfield  {journal} {\bibinfo  {journal}
  {Sov. Phys. Usp.}\ }\textbf {\bibinfo {volume} {13}},\ \bibinfo {pages} {408}
  (\bibinfo {year} {1970}{\natexlab{b}})}\BibitemShut {NoStop}%
\bibitem [{\citenamefont {Bean}\ and\ \citenamefont
  {Livingston}(1964)}]{Bean64}%
  \BibitemOpen
  \bibfield  {author} {\bibinfo {author} {\bibfnamefont {C.~P.}\ \bibnamefont
  {Bean}}\ and\ \bibinfo {author} {\bibfnamefont {J.~D.}\ \bibnamefont
  {Livingston}},\ }\bibfield  {title} {\bibinfo {title} {{Surface Barrier in
  Type-II Superconductors}},\ }\href@noop {} {\bibfield  {journal} {\bibinfo
  {journal} {Phys. Rev. Lett.}\ }\textbf {\bibinfo {volume} {12}},\ \bibinfo
  {pages} {14} (\bibinfo {year} {1964})}\BibitemShut {NoStop}%
\bibitem [{\citenamefont {Fossheim}\ and\ \citenamefont
  {Sudboe}(2004)}]{FossheimBook}%
  \BibitemOpen
  \bibfield  {author} {\bibinfo {author} {\bibfnamefont {K.}~\bibnamefont
  {Fossheim}}\ and\ \bibinfo {author} {\bibfnamefont {A.}~\bibnamefont
  {Sudboe}},\ }\href@noop {} {\emph {\bibinfo {title} {{Superconductivity:
  Physics and Applications}}}}\ (\bibinfo  {publisher} {Wiley},\ \bibinfo
  {year} {2004})\ p.\ \bibinfo {pages} {232}\BibitemShut {NoStop}%
\bibitem [{\citenamefont {Lyu}\ \emph {et~al.}(2021)\citenamefont {Lyu},
  \citenamefont {Jiang}, \citenamefont {Wang}, \citenamefont {Xiao},
  \citenamefont {Dong}, \citenamefont {Chen}, \citenamefont
  {Milo{\v{s}}evi{\'{c}}}, \citenamefont {Wang}, \citenamefont {Divan},
  \citenamefont {Pearson}, \citenamefont {Wu}, \citenamefont {Peeters},\ and\
  \citenamefont {Kwok}}]{Lyu21}%
  \BibitemOpen
  \bibfield  {author} {\bibinfo {author} {\bibfnamefont {Y.-Y.}\ \bibnamefont
  {Lyu}}, \bibinfo {author} {\bibfnamefont {J.}~\bibnamefont {Jiang}}, \bibinfo
  {author} {\bibfnamefont {Y.-L.}\ \bibnamefont {Wang}}, \bibinfo {author}
  {\bibfnamefont {Z.-L.}\ \bibnamefont {Xiao}}, \bibinfo {author}
  {\bibfnamefont {S.}~\bibnamefont {Dong}}, \bibinfo {author} {\bibfnamefont
  {Q.-H.}\ \bibnamefont {Chen}}, \bibinfo {author} {\bibfnamefont {M.~V.}\
  \bibnamefont {Milo{\v{s}}evi{\'{c}}}}, \bibinfo {author} {\bibfnamefont
  {H.}~\bibnamefont {Wang}}, \bibinfo {author} {\bibfnamefont {R.}~\bibnamefont
  {Divan}}, \bibinfo {author} {\bibfnamefont {J.~E.}\ \bibnamefont {Pearson}},
  \bibinfo {author} {\bibfnamefont {P.}~\bibnamefont {Wu}}, \bibinfo {author}
  {\bibfnamefont {F.~M.}\ \bibnamefont {Peeters}},\ and\ \bibinfo {author}
  {\bibfnamefont {W.-K.}\ \bibnamefont {Kwok}},\ }\bibfield  {title} {\bibinfo
  {title} {Superconducting diode effect via conformal-mapped nanoholes},\
  }\href@noop {} {\bibfield  {journal} {\bibinfo  {journal} {Nature
  Communications}\ }\textbf {\bibinfo {volume} {12}},\ \bibinfo {pages} {2703}
  (\bibinfo {year} {2021})}\BibitemShut {NoStop}%
\bibitem [{\citenamefont {Sundahl}(2019)}]{SundahlThesis}%
  \BibitemOpen
  \bibfield  {author} {\bibinfo {author} {\bibfnamefont {C.~S.}\ \bibnamefont
  {Sundahl}},\ }\emph {\bibinfo {title} {{Synthesis of Superconducting Nb$_3$Sn
  Thin Film Heterostructures for the Study of High-Energy RF Physics}}},\
  \href@noop {} {Ph.D. thesis},\ \bibinfo  {school} {The University of
  Wisconsin - Madison}, \bibinfo {address} {ProQuest Dissertations Publishing,
  13805402} (\bibinfo {year} {2019})\BibitemShut {NoStop}%
\bibitem [{\citenamefont {Charlesworth}\ \emph {et~al.}(1970)\citenamefont
  {Charlesworth}, \citenamefont {Macphail},\ and\ \citenamefont
  {Madsen}}]{Charlesworth70}%
  \BibitemOpen
  \bibfield  {author} {\bibinfo {author} {\bibfnamefont {J.~P.}\ \bibnamefont
  {Charlesworth}}, \bibinfo {author} {\bibfnamefont {I.}~\bibnamefont
  {Macphail}},\ and\ \bibinfo {author} {\bibfnamefont {P.~E.}\ \bibnamefont
  {Madsen}},\ }\bibfield  {title} {\bibinfo {title} {Experimental work on the
  niobium-tin constitution diagram and related studies},\ }\href@noop {}
  {\bibfield  {journal} {\bibinfo  {journal} {Journal of Materials Science}\
  }\textbf {\bibinfo {volume} {5}},\ \bibinfo {pages} {580} (\bibinfo {year}
  {1970})}\BibitemShut {NoStop}%
\bibitem [{\citenamefont {Berdiyorov}\ \emph {et~al.}(2009)\citenamefont
  {Berdiyorov}, \citenamefont {Elmurodov}, \citenamefont {Peeters},\ and\
  \citenamefont {Vodolazov}}]{Berdiyorov09}%
  \BibitemOpen
  \bibfield  {author} {\bibinfo {author} {\bibfnamefont {G.~R.}\ \bibnamefont
  {Berdiyorov}}, \bibinfo {author} {\bibfnamefont {A.~K.}\ \bibnamefont
  {Elmurodov}}, \bibinfo {author} {\bibfnamefont {F.~M.}\ \bibnamefont
  {Peeters}},\ and\ \bibinfo {author} {\bibfnamefont {D.~Y.}\ \bibnamefont
  {Vodolazov}},\ }\bibfield  {title} {\bibinfo {title} {{Finite-size effect on
  the resistive state in a mesoscopic type-II superconducting stripe}},\
  }\href@noop {} {\bibfield  {journal} {\bibinfo  {journal} {Phys. Rev. B}\
  }\textbf {\bibinfo {volume} {79}},\ \bibinfo {pages} {174506} (\bibinfo
  {year} {2009})}\BibitemShut {NoStop}%
\bibitem [{\citenamefont {Sánchez-Lotero}\ \emph {et~al.}(2014)\citenamefont
  {Sánchez-Lotero}, \citenamefont {{Albino Aguiar}},\ and\ \citenamefont
  {Domínguez}}]{Lotero14}%
  \BibitemOpen
  \bibfield  {author} {\bibinfo {author} {\bibfnamefont {P.}~\bibnamefont
  {Sánchez-Lotero}}, \bibinfo {author} {\bibfnamefont {J.}~\bibnamefont
  {{Albino Aguiar}}},\ and\ \bibinfo {author} {\bibfnamefont {D.}~\bibnamefont
  {Domínguez}},\ }\bibfield  {title} {\bibinfo {title} {Behavior of the
  flux-flow resistivity in mesoscopic superconductors},\ }\href@noop {}
  {\bibfield  {journal} {\bibinfo  {journal} {Physica C: Superconductivity and
  its Applications}\ }\textbf {\bibinfo {volume} {503}},\ \bibinfo {pages}
  {120} (\bibinfo {year} {2014})}\BibitemShut {NoStop}%
\bibitem [{\citenamefont {Vodolazov}\ and\ \citenamefont
  {Peeters}(2007)}]{Vodolazov07}%
  \BibitemOpen
  \bibfield  {author} {\bibinfo {author} {\bibfnamefont {D.~Y.}\ \bibnamefont
  {Vodolazov}}\ and\ \bibinfo {author} {\bibfnamefont {F.~M.}\ \bibnamefont
  {Peeters}},\ }\bibfield  {title} {\bibinfo {title} {Rearrangement of the
  vortex lattice due to instabilities of vortex flow},\ }\href@noop {}
  {\bibfield  {journal} {\bibinfo  {journal} {Phys. Rev. B}\ }\textbf {\bibinfo
  {volume} {76}},\ \bibinfo {pages} {014521} (\bibinfo {year}
  {2007})}\BibitemShut {NoStop}%
\bibitem [{\citenamefont {Gulian}(2020)}]{GulianCOMSOL}%
  \BibitemOpen
  \bibfield  {author} {\bibinfo {author} {\bibfnamefont {A.}~\bibnamefont
  {Gulian}},\ }\href@noop {} {\emph {\bibinfo {title} {{S}hortcut to
  {S}uperconductivity: {S}uperconducting {E}lectronics via {COMSOL}
  {M}odeling}}}\ (\bibinfo  {publisher} {Springer},\ \bibinfo {year}
  {2020})\BibitemShut {NoStop}%
\bibitem [{\citenamefont {Weber}\ and\ \citenamefont {Kramer}(1991)}]{Weber91}%
  \BibitemOpen
  \bibfield  {author} {\bibinfo {author} {\bibfnamefont {A.}~\bibnamefont
  {Weber}}\ and\ \bibinfo {author} {\bibfnamefont {L.}~\bibnamefont {Kramer}},\
  }\bibfield  {title} {\bibinfo {title} {Dissipative states in a
  current-carrying superconducting film},\ }\href@noop {} {\bibfield  {journal}
  {\bibinfo  {journal} {Journal of Low Temperature Physics}\ }\textbf {\bibinfo
  {volume} {84}},\ \bibinfo {pages} {289} (\bibinfo {year} {1991})}\BibitemShut
  {NoStop}%
\bibitem [{\citenamefont {Andronov}\ \emph {et~al.}(1993)\citenamefont
  {Andronov}, \citenamefont {Gordion}, \citenamefont {Kurin}, \citenamefont
  {Nefedov},\ and\ \citenamefont {Shereshevsky}}]{Andronov93}%
  \BibitemOpen
  \bibfield  {author} {\bibinfo {author} {\bibfnamefont {A.}~\bibnamefont
  {Andronov}}, \bibinfo {author} {\bibfnamefont {I.}~\bibnamefont {Gordion}},
  \bibinfo {author} {\bibfnamefont {V.}~\bibnamefont {Kurin}}, \bibinfo
  {author} {\bibfnamefont {I.}~\bibnamefont {Nefedov}},\ and\ \bibinfo {author}
  {\bibfnamefont {I.}~\bibnamefont {Shereshevsky}},\ }\bibfield  {title}
  {\bibinfo {title} {Kinematic vortices and phase slip lines in the dynamics of
  the resistive state of narrow superconductive thin film channels},\
  }\href@noop {} {\bibfield  {journal} {\bibinfo  {journal} {Physica C:
  Superconductivity and its Applications}\ }\textbf {\bibinfo {volume} {213}},\
  \bibinfo {pages} {193} (\bibinfo {year} {1993})}\BibitemShut {NoStop}%
\end{thebibliography}

\providecommand{\noopsort}[1]{}\providecommand{\singleletter}[1]{#1}%

\end{document}